\documentclass[aps,pra,nofootinbib,preprint,
showpacs,amsmath,amssymb,floatfix,showpacs]{revtex4}
\usepackage{amssymb}
\usepackage{amsmath}
\usepackage{bm}
\usepackage{upgreek}
\usepackage{graphicx}

\begin{document}

\title{Minkowski momentum resulting from a vacuum-medium mapping procedure, and a brief review of Minkowski momentum experiments}         % Enter your title between curly braces

\author{Iver Brevik}
\address{Department of Energy and Process Engineering, Norwegian University of Science and Technology, 7491 Trondheim, Norway}

\date{\today}          % Enter your date or \today between curly braces
%\maketitle

\begin{abstract}

A discussion is given on the interpretation and physical importance of the Minkowski momentum in macroscopic electrodynamics (essential for the Abraham-Minkowski problem). We focus on the following two facets: (1)  Adopting a simple dielectric model where the refractive index $n$ is constant, we demonstrate by means of a mapping procedure how the electromagnetic field in a medium can be mapped into a corresponding  field in vacuum. This mapping was  presented many years ago [I. Brevik and B. Lautrup, Mat. Fys. Medd. Dan. Vid. Selsk {\bf 38}(1), 1 (1970)], but is apparently not well known. A characteristic property of this procedure  is that it shows how natural the Minkowski energy-momentum tensor fits into the canonical formalism. Especially the spacelike character of the electromagnetic total four-momentum  for a radiation field (implying negative electromagnetic energy in some inertial frames), so strikingly demonstrated in the Cherenkov effect, is worth attention. (2) Our second objective is to give a critical analysis of some recent experiments on electromagnetic momentum.  Care must here be taken in the interpretations: it is easy to be misled and conclude that an experiment is important for the energy-momentum problem, while what is demonstrated experimentally is merely the action of the Abraham-Minkowski force acting in  surface layers or inhomogeneous regions. The Abraham-Minkowski force is common for the two energy-momentum tensors and carries no information about field momentum. As a final item,  we propose an experiment that might show the existence of the Abraham force at high frequencies. This would eventually be a welcome  optical analogue to the classic low-frequency 1975 Lahoz-Walker experiment.

\end{abstract}
\maketitle

\bigskip
\section{Introduction} \label{sec1}
It is almost amusing to note that the Abraham-Minkowski energy-momentum problem in dielectric media - considered to be a very old-fashioned problem in the 1960's when the present author began working on it - has in recent years emerged to become of considerable interest. Some recent papers on the theme  are listed in Refs.~\cite{pfeifer07,hinds09,barnett10,baxter10,barnett10a,mansuripur10,milonni10,kemp11,rikken11,griffiths12,testa13,mansuripur13,astrath14,leonhardt14,
barnett15,zhang15,kemp15,aanensen13,sheppard16,wang16,nesterenko16,nesterenko16a}. The author published   three papers at the Royal Danish Academy of Sciences in 1970 \cite{brevik70,brevik70a,breviklautrup70} (the last one on QED with Benny Lautrup). Some years later   a review was published in 1979 \cite{brevik79}, dealing with  experimental consequences of the theory. Among more recent activities of the author within the same area we  mention a couple of papers \cite{brevik10,brevik11}, the first of which focusing again on experimental possibilities, namely how to use intensity-modulated whispering gallery modes in a microresonator to measure the Abraham force at optical frequencies.

The reason why we are revisiting this topic here, is twofold:

\noindent 1. We wish to re-emphasize that there is an intimate connection between the Maxwell field equations in vacuum and in an isotropic nondispersive; the two cases can be related by a simple transformation procedure.  In turn, this mapping leads on to the Minkowski momentum in a straightforward way. Actually this procedure was spelled out already in  1970  \cite{breviklautrup70}, but has apparently not been much recognized in the literature. In essence, the method demonstrates how natural the Minkowski theory fits into the canonical procedure in field theory. On a deeper level, this is a consequence of the vanishing of the Minkowski four-force for a pure radiation field in matter, implying that the  total field energy and momentum make up a four-vector \cite{moller72}. The fact that the Minkowski four-momentum turns out to be spacelike instead of timelike as is usually the case, does not disturb this fundamental property.

\noindent 2. Formal arguments of the above kind, although impressive and elegant, of course cannot determine whether the Minkowski theory is physically the correct one  (strictly speaking this is a matter of convenience rather than correctness). One ought to  go to real experiments to get information. And this brings us to the second theme of this paper, which is to make a brief critical survey over recent experiments in optics, and judge to what extent they elucidate the photon momentum problem.

We shall be concerned with the simple case of an isotropic nondispersive medium moving at constant four-velocity $V_\mu$ in the laboratory system. We begin in the next section by presenting the transformation matrix $b_{\mu\nu}$, and show how this can be used to map the Maxwell equations in vacuum to those in a medium in the case of a pure radiation field. In Sec. III we include extraneous charges and currents, and show how the vacuum potential $A_\mu^{\rm vac}$ at spacetime position $x_\mu$ can be mapped into a medium potential $B_\mu$ at another spacetime position $y_\mu=b_{\mu\nu}x^\nu$. The gauge condition is also considered. In Sec. IV we quantize the electromagnetic field by means of the transformation procedure. At this point it will turn out how the canonical total four-momentum becomes naturally identifiable with the Minkowski four-momentum.

Section V contains the mentioned discussion about some recent experiments. Care should be taken to see whether the experiments say something about electromagnetic momentum, or whether they merely show the action of the force ${\bf f}^{\rm AM}$ which act in the air-liquid boundary (cf. Eq.~(\ref{58}) below). This force is common for the Abraham and Minkowski tensors.

We point out that we do not intend to give an exhaustive review of the developments within macroscopic electrodynamics in media. Historically, the papers of Jauch and Watson \cite{jauch48} play an important role. And there are  several more recent papers in this area; cf., for instance, Refs.~\cite{huttner92,philbin10,braun11}.

\section{The transformation matrix. Pure radiation field}

For convenience we adopt the same conventions as in Ref.~\cite{breviklautrup70}. Thus the coordinates are $x_\mu=(x_0,x_1,x_2,x_3)=(t, {\bf x})$, the metric tensor $g_{\mu\nu}$ has diagonal components $(1,-1,-1, -1)$, and we put $\hbar=c=1$. We adopt a very simple material  model where the medium is isotropic and nondispersive, moving in the laboratory inertial system with constant four-velocity $V_\mu=(V_0, {\bf V})=\gamma (1, {\bf v})$, where $\gamma=(1-v^2)^{-1/2}$, $V^2=V^\mu V_\mu=1$.

\subsection{Configuration space}

There are two field tensors, $F_{\mu\nu}$ and $H_{\mu\nu}$, defined by $F_{i0}=E_i, \ F_{ij}=-B_k$ (cycl), $H_{i0}=D_i, \ H_{ij}=-H_k$ (cycl). This means that the Maxwell equations for the pure radiation field can be expressed covariantly, in any inertial system, as
\begin{equation}
\partial_\lambda F_{\mu\nu}+\partial_\mu F_{\nu\lambda}+\partial_\nu F_{\lambda \mu}=0, \label{1}
\end{equation}
\begin{equation}
\partial^\nu H_{\mu\nu}=0, \label{2}
\end{equation}
where $\partial^\nu=\partial/\partial x^\nu$. The covariant constitutive relation can be expressed as
\begin{equation}
\mu H_{\alpha\beta}=F_{\alpha\beta}+\kappa (F_\alpha V_\beta -F_\beta V_\alpha), \label{3}
\end{equation}
where we have defined
\begin{equation}
\kappa=\varepsilon\mu-1=n^2-1, \quad F_\alpha=F_{\alpha\beta}V^\beta, \label{4}
\end{equation}
($n=\sqrt{\varepsilon\mu}$ is thus the refractive index in the medium's rest system)

We can now introduce the transformation matrix $b_{\mu\nu}$,
\begin{equation}
b_{\mu\nu}=g_{\mu\nu}+(n-1)V_\mu V_\nu. \label{5}
\end{equation}
It satisfies the convenient property
\begin{equation}
(b^p)_{\mu\nu}=g_{\mu\nu}+(n^p-1)V_\mu V_\nu, \label{6}
\end{equation}
valid for all integers $p$, positive and negative. This enables us to write the constitutive relation (\ref{3}) in compact form as
\begin{equation}
\mu H_{\alpha\beta}=(b^2)_\alpha^\rho (b^2)_\beta^\sigma F_{\rho\sigma}. \label{7}
\end{equation}
We now introduce the electromagnetic potentials via $F_{\mu\nu}=\partial_\mu A_\nu-\partial_\nu A_\mu$, and consider the Lorentz gauge
\begin{equation}
\Lambda^{\rm F}(x) \equiv (b^2)^{\mu\nu}\partial_\mu A_\nu=\partial \cdot A+\kappa \partial \cdot V A\cdot V=0, \label{8}
\end{equation}
with $\partial \cdot A=\partial_\mu A^\mu$. Making use of Eq.~(\ref{7}) we see that the field equation (\ref{2}) takes the elegant form
\begin{equation}
(b^2)^{\rho\sigma}\partial_\rho\partial_\sigma A_\mu=[\Box +\kappa(\partial \cdot V)^2]A_\mu=0, \label{9}
\end{equation}
where $\Box=\partial_\mu \partial^\mu$. Equation (\ref{1}) is automatically satisfied.

We now note  that the field equation (\ref{9}) can also be obtained from a variational principle, corresponding to the Lagrangian density
\begin{equation}
L=-\frac{1}{4}F_{\mu\nu}H^{\mu\nu}-\frac{1}{2\mu}(\Lambda^{\rm F})^2, \label{10}
\end{equation}
although one should here be aware of the restriction that the variational equation is the same as the Maxwell field equation (\ref{2}) only if the subsidiary condition $\Lambda^{\rm F}=0$ is imposed explicitly.

The expression (\ref{10}) is the starting point for canonical quantization of the radiation field. The canonically conjugate momenta are
\begin{equation}
\pi^\mu=H^{\mu 0}-\frac{1}{\mu}(b^2)^{\mu 0} \Lambda^{\rm F}. \label{11}
\end{equation}
The canonical commutation rules become now formally the conventional ones. We give here only the nontrivial one,
\begin{equation}
[\pi^\mu(x), A_\nu(x^\prime)]_{x_0=x_0^{\prime}}=-ig_\nu^\mu \delta(\bf x-x^{\prime}), \label{12}
\end{equation}
the other ones vanishing.

Consider next the relativistic invariance of the quantization procedure. For a  physical system that is {\it closed}, the invariance is usually checked by verifying that the operators $P_\mu$ for total four-momentum and $M_{\mu\nu}$ for total angular momentum are constants of motion, and moreover identifiable with   the Hermitian operator ${\mathcal{P}}_{\mu}$ generating infinitesimal translations and the corresponding operator ${\mathcal{M}}_{\mu\nu}$ generating infinitesimal rotations in four-space. These operators satisfy
\begin{equation}
i[{\mathcal{P}}_\mu, A_\nu (x)]=\partial_\mu A_\nu (x), \label{13}
\end{equation}
\begin{equation}
i[ {\mathcal{M}}_{\mu\nu}, A^\sigma(x)]=x_\mu \partial_\nu A^\sigma (x)-x_\nu \partial_\mu A^\sigma (x)+I_{\mu\nu}^{\sigma\rho} A_\rho (x), \label{14}
\end{equation}
where $I_{\mu\nu}^{\rho \sigma}=g_\mu^\sigma g_\nu^\rho-g_\mu^\rho g_\nu^\sigma$.

It is now possible to verify that Eq.~(\ref{13}) is valid also if ${\mathcal{P}}_\mu$ becomes replaced by the field operator $P_\mu$. To this end we may  start from the canonical energy-momentum tensor
\begin{equation}
S_{\mu\nu}^{\rm can}=-g_{\mu\nu}L+\frac{\partial L}{\partial \partial^\nu A_\alpha}\partial_\mu A_\alpha, \label{15}
\end{equation}
from which $P_\mu$ follows by integration over the volume,
\begin{equation}
P_\mu=\int S_{\mu 0}^{\rm can} dV. \label{16}
\end{equation}
In view of the field equations, $P_\mu$ is a constant, and so we find by using  Eqs.~(\ref{15}) and (\ref{16}) that  Eq.~(\ref{13}) is valid if ${\mathcal{P}}_\mu$ is replaced by $P_\mu$.

The case of angular momentum becomes more complicated to analyze, there occur formal  ambiguities related to the fact that we are dealing with a non-closed physical system: the Lagrangian (\ref {10}) describes the radiation field and its interaction with the medium but not the medium itself. It implies here that the angular momentum $M_{\mu\nu}$ is not a constant of
 motion. A more detailed discussion on this case can be found in Ref.~\cite{breviklautrup70}. We restrict ourselves here to stating that in the classical field theory where one obtains correspondence with Maxwell theory simply by setting $\Lambda^{\rm F}=0$, also the field angular momentum $M_{\mu\nu}$ can be substituted for the Hermitian operator ${\mathcal{M}}_{\mu\nu}$ in Eq.~(\ref{14}), assumed that $M_{\mu\nu}$ is calculated with use of Minkowski's energy-momentum tensor (superscript M),
\begin{equation}
S_{\mu\nu}^{\rm M}= -F_{\mu\alpha}H_\nu^\alpha +\frac{1}{4}g_{\mu\nu}F_{\alpha\beta} H^{\alpha\beta}. \label{17}
\end{equation}

\subsection{Momentum space}

It is of interest to consider the field equation in momentum space.  We may start from the integral
\begin{equation}
A_\mu(x)=(2\pi)^{-3/2}\int dk \delta (k^2+\kappa (k\cdot V)^2)e^{-ikx}A_\mu(k), \label{18}
\end{equation}
where $k^2=k^\mu k_\mu$ and $dk=dk_0d\bf k$. The field equation (\ref{9})  in configuration space then leads to
\begin{equation}
k^2+\kappa (k \cdot  V)^2=0. \label{19}
\end{equation}
For the angular  frequency $k_0$ there are two solutions, $k_a$ and $k_b$, where
\begin{equation}
k_{a,b}=\frac{\kappa V_0\,{\bf k \cdot V} \pm \sqrt{(1+\kappa V_0^2){\bf k}^2-\kappa {\bf (k\cdot V)}^2}}{1+\kappa V_0^2}. \label{19}
\end{equation}
It should here be noticed that there exist some inertial systems in which $k_a$ (upper sign) becomes negative. This is when $\kappa {\bf V}^2>1$, i.e. $n^2v^2>1$, and corresponds to the Cherenkov effect. Evidently this is a consequence of the fact that the Minkowski four-momentum  $P_\mu^{\rm M}$ of a radiation field is spacelike. It can immediately by visualized by considering a Cherenkov emitter in the inertial system where it is at rest prior to the emission: the emitter sending out a photon within the Cherenkov cone gets a kick and receives thus an increase of kinetic energy. The energy of the photon itself accordingly has to be  negative due to the energy balance.

It is also instructive to use Eq.~(\ref{18}) to write down the equation for the surface $k_0=~$const in $\bf k-$space. With the coordinate axes oriented such that $V_1=|{\bf V}|, ~V_2=V_3=0$, we get
\begin{equation}
\frac{[k_1+\kappa k_0V_0|{\bf V}|(1-\kappa {\bf V}^2)^{-1}]^2}{n^2k_0^2(1-\kappa {\bf V}^2)^{-2}}+\frac{k_2^2}{n^2k_0^2(1-\kappa {\bf V}^2)^{-1}}
+\frac{k_3^2}{n^2k_0^2(1-\kappa {\bf V}^2)^{-1}} =0.\label{20}
\end{equation}
If $\kappa {\bf V}^2<1$, this is the equation of an ellipsoid in ${\bf k-}$space. If  $\kappa {\bf V}^2>1$, it is the equation of a two-sheet hyperboloid such that one sheet corresponds to the solution $k_0=k_a({\bf k})$ and the other sheet corresponds to $k_0=k_b({\bf k})$.

 Let us make a remark about the commutation rules: if we write the delta function in Eq.~(\ref{18}) as a sum of to delta functions containing $(k_0-k_a)$ and $(k_0-k_b)$, and thereafter integrate over $k_0$, we obtain the following expansion for the potential,
\begin{equation}
A_\mu(x)=(2\pi)^{-3/2}\int d{\bf k}\left[ \frac{\mu}{(1+\kappa V_0^2)(k_a-k_b)}\right]^{1/2}(b^{-1})_\mu^\nu(e^{-ik\cdot x}a_\nu({\bf k})+e^{ik\cdot x}a_\nu^\dagger({\bf k})), \label{21}
\end{equation}
from which we obtain the usual commutation rules for the operators $a_\mu$,
\begin{equation}
[a_\mu({\bf k}), a_\nu^\dagger ({\bf k}^\prime]=-g_{\mu\nu}\delta(\bf{k}-{\bf k}^\prime), \label{22}
\end{equation}
the other commutators vanishing.

From Eqs.~(\ref{21}) and (\ref{22}) we obtain the commutation rules,
\begin{equation}
[A_\mu (x), A_\nu (x^\prime)] =-\frac{i\mu}{n}(b^{-2})_{\mu\nu}D^{\rm M}(x-x^\prime), \label{A}
\end{equation}
where $D^{\rm M}$ is the invariant singular function according to Minkowski,
\begin{equation}
D^{\rm M }(x)=-\frac{in}{(2\pi)^3}\int dk e^{-ik\cdot x}\,\delta (k^2+\kappa (k\cdot V)^2)\varepsilon (k\cdot V). \label{B}
\end{equation}
Here $\varepsilon (x)$ is the step function, $\varepsilon(x)=1$ if $x>0$ and $\varepsilon(x)=-1$ if $x<0$.

Finally, the total canonical four-momentum defined by Eqs.~(\ref{15}) and (\ref{16}) can be expressed in Fourier space, after inserting the expression (\ref{10}) for $L$. Some calculation yields
\begin{equation}
P_\mu=-\frac{1}{2}\int d{\bf k}k_\mu \{a_\nu({\bf k}), a^{\nu \dagger}(\bf{k})\}, \label{23}
\end{equation}
the curly bracket meaning the anticommutator.

\section{Vacuum-medium mapping of the electromagnetic field}

We now admit the presence of extraneous charges and currents in the moving medium, so that the Maxwell equations (\ref{2}) become replaced by
\begin{equation}
\partial^\nu H_{\mu\nu}=-j_\mu, \label{25}
\end{equation}
where $j_\mu=(\rho, {\bf j})$. We will show how one can start from the known expressions in vacuum electrodynamics and herefrom derive the corresponding equation in medium electrodynamics by applying the transformation procedure above.

\subsection{Classical theory}

We establish the vacuum-medium mapping in two steps. First, introduce new $B-$potentials,
\begin{equation}
B_\mu (x)=b_{\mu\nu}A^\nu (x), \label{26}
\end{equation}
as well as a new differential operator,
\begin{equation}
D_\mu=b_{\mu\nu}\partial^\nu. \label{27}
\end{equation}
We can then associate the $B-$fields with field strengths, called $G_{\mu\nu}$,
\begin{equation}
G_{\mu\nu}(x)=b_\mu^\rho b_\nu^\sigma F_{\rho\sigma}(x)=D_\mu B_\nu (x)-D_\nu B_\mu(x). \label{28}
\end{equation}
This implies that we can express the field equations (\ref{25}) in the form
\begin{equation}
D^\nu G_{\mu\nu}=-D^2 B_\mu+D_\mu D_\nu B^\nu=-J_\mu, \label{29}
\end{equation}
where $J_\mu=\mu (b^{-1})_\mu^\nu j_\nu$ is the current four-vector density of the $B-$field. It satisfies the continuity equation
\begin{equation}
D^\mu J_\mu=0. \label{30}
\end{equation}
It is apparent that the analogy with vacuum electrodynamics is very close. Our second step in the mapping is to define the vacuum potentials as
\begin{equation}
A_\mu^{\rm vac} (x)=\rho B_\mu(y)=\rho b_{\mu\nu}A^\nu (x), \label{31}
\end{equation}
where
\begin{equation}
y_\mu=b_{\mu\nu}x^\nu,   \label{32}
\end{equation}
\begin{equation}
\rho=(n/\mu)^{1/2}. \label{33}
\end{equation}
Equation (\ref{32}) implies that $D_\mu^y$ can be replaced by $\partial_\mu^ x$, and so the vacuum equations for $A_\mu^{\rm vac}(x)$ take the form
\begin{equation}
-\Box  A_\mu^{\rm vac}(x)+\partial_\mu\partial^\nu A_\nu^{\rm vac}(x)=-j_\mu^{\rm vac}(x), \label{34}
\end{equation}
where the current density in vacuum is
\begin{equation}
j_\mu^{\rm vac}(x)=\rho J_\mu(y)=\rho \mu (b^{-1})_\mu^\nu j_\nu(y), \label{35}
\end{equation}
satisfying the continuity equation
\begin{equation}
\partial^\mu j_\mu^{\rm vac}(x)=0. \label{36}
\end{equation}
Finally, we make some brief remarks on polarization vectors. Let $l_\mu$ be the four-momentum of a photon in vacuum. We write the one-photon potential as
\begin{equation}
A_\mu^{\rm vac}(x)=e^{-il\cdot x}e_\mu({\bf l})+e^{il\cdot x}e_\mu^*(\bf{l}), \label{37}
\end{equation}
where the polarization four-vector $e_\mu$ satisfies the normalization condition $e_\mu^* e^\mu=-1$. The one-photon potential in the medium is then
\begin{equation}
A_\mu(x)=\frac{1}{\rho}(e^{-ik\cdot x}f_\mu+e^{ik\cdot x}f_\mu^*), \label{38}
\end{equation}
with $k_\mu$ the corresponding four-momentum. From Eqs.~(\ref{31}) and (\ref{32}) it follows that
\begin{equation}
k_\mu=(b^{-1})_\mu^\nu \,l_\nu, \label{39}
\end{equation}
and the corresponding relation for the polarization four-vectors is
\begin{equation}
f_\mu=(b^{-1})_\mu^\nu \, e_\nu. \label{40}
\end{equation}
We orient the coordinate axes such that ${\bf{e}}^{(1)}$ is collinear with
$\bf{l}$ in the rest frame of the medium,
\begin{equation}
{\bf{e}}^{(2)}\times {\bf{e}}^{(3)} =
{\bf l}/{|\bf{l}|}= {\bf k}/{|\bf{k}}|. \label{41}
\end{equation}
This implies that the covariant polarization sum in the vacuum case,
\begin{equation}
\sum_{\lambda=2}^3 e_\mu^{(\lambda)}e_\nu^{(\lambda)*}=-g_{\mu\nu}-\frac{l_\mu l_\nu}{(l\cdot V)^2}+\frac{l_\mu V_\nu+l_\nu V_\mu}{l\cdot V}, \label{42}
\end{equation}
can for the medium be expressed in terms of $f_\mu$ and $k_\mu$ as
\begin{equation}
\sum_{\lambda=2}^3 f_\mu^{(\lambda)}f_\nu^{(\lambda)*}= -(b^{-2})_{\mu\nu}- \frac{k_\mu k_\nu}{(k\cdot V)^2}+\frac{k_\mu V_\nu+k_\nu V_\mu}{k\cdot V}. \label{43}
\end{equation}

\section{Quantization via the mapping method}

We proceed to the quantum theory, starting from the theory in vacuum and making use of the mapping technique. The theory can readily be formulated in covariant gauges, along the lines presented earlier by one of us in the vacuum case ($\kappa =0$) \cite{lautrup67}, but for simplicity we will restrict ourselves to the simple case of the Fermi gauge here.

We start from the Lagrangian density
\begin{equation}
L^{\rm vac}(x)=-\frac{1}{4}F_{\mu\nu}^{\rm vac} F^{\mu\nu \rm ~vac}-
\Lambda^{\rm vac}\partial^\mu A_\mu^{\rm vac}+
\frac{1}{2}(\Lambda^{\rm vac})^2, \label{44}
\end{equation}
The field equations obtained from Eq.~(\ref{44}) are
\begin{equation}
-\Box A_\mu^{\rm vac}+\partial_\mu\partial^\nu A_\nu^{\rm vac}=\partial_\mu \Lambda^{\rm vac}. \label{45}
\end{equation}
By varying with respect to $\Lambda^{\rm vac}$ we obtain the gauge condition,
\begin{equation}
\partial^\mu A_\mu^{\rm vac}=\Lambda^{\rm vac}. \label{46}
\end{equation}
In case of the Fermi gauge the commutations become simple,
\begin{equation}
[A_\mu^{\rm vac}(x), A_\nu^{\rm vac}(x^\prime)]=-ig_{\mu\nu}D(x-x^\prime), \label{47}
\end{equation}
where $D(x)$ is the singular function
\begin{equation}
D(x)=-\frac{i}{(2\pi)^3}\int dl \varepsilon (l)\delta(l^2)e^{-il\cdot x}=-\frac{\varepsilon (x)}{2\pi}\delta(x^2). \label{48}
\end{equation}
We can now construct the fields $B_\mu(y)$ and $A_\mu(y)$ on the basis of $A_\mu^{\rm vac}(x)$, using Eqs.~(\ref{31}) - (\ref{33}). For the $B-$field the field equations and the gauge condition are
\begin{equation}
-D^2B_\mu(y)+D_\mu D^\nu B_\nu(y)=D_\mu \Lambda (y), \label{49}
\end{equation}
\begin{equation}
D^\mu B_\mu(y)=\Lambda (y), \label{50}
\end{equation}
where $\Lambda(y)=\rho^{-1}\Lambda^{\rm vac}(x)$. For the $B-$field the commutation rules become
\begin{equation}
[B_\mu(y), B_\nu(y^\prime)]=-\frac{i}{\rho^2}g_{\mu\nu}D(b^{-1}(y-y^\prime)), \label{51}
\end{equation}
where $(b^{-1}y)_\mu=x_\mu$, and for the $A-$ field we obtain
\begin{equation}
[A_\mu(y), A_\nu(y^\prime)]=-\frac{i}{\rho^2}(b^{-2})_{\mu\nu}D(b^{-1}(y-y^\prime)). \label{52}
\end{equation}
The singular function in the last expression can be rewritten in a more conventional form as
\begin{equation}
D(b^{-1}y)=-\frac{in}{(2\pi)^3}\int dk \varepsilon (k\cdot V)\delta (k^2+\kappa (k\cdot V)^2)e^{-ik\cdot y}. \label{53}
\end{equation}It is thus apparent that $D(b^{-1}y)$ is equal to $D^{\rm M}(y)$ defined in Eq.~(\ref{8}), and the commutation relations (\ref{52}) and (\ref{A}) are the same.

Consider now the four-momentum $P^\mu$ for the free radiation field. In the vacuum case we have
\begin{equation}
i[P_\mu^{\rm vac}, A_\nu^{\rm vac}(x)]=\partial_\mu A_\nu^{\rm vac}(x), \label{54}
\end{equation}
cf. the operator equation (\ref{13}). Thus, the quantity
\begin{equation}
P_\mu=(b^{-1})_\mu^\nu P_\nu^{\rm vac} \label{55}
\end{equation}
satisfies
\begin{equation}
i[P_\mu, A_\nu (y)]=\partial_\mu^y A_\nu (y),\label{56}
\end{equation}
and is to be interpreted as the Minkowski four-momentum of the radiation field.

\section{Analysis of some recent experimental results, and a proposal}

In the previous section our purpose was to emphasize how well the Minkowski tensor adapts itself to the canonical formalism in classical and quantum mechanical field theory. The usefulness of this tensor for explaining real situations can however not be decided definitely upon from these formal arguments. We need additional information, especially from the experimental side. In this section we use dimensional units. Our comments in the following will be brief.

Let us first state the following: all the experiments that we are aware of in optics, are explainable in terms of the Minkowski tensor in a straightforward way. It is here worth noticing that this tensor is after all not corresponding to the physical force density $\bf f$ which one derives from quite simple arguments -  cf., for instance, Refs.~\cite{brevik79,ginzburg89,landau84},
\begin{equation}
{\bf f}= {\bf f}^{\rm AM}+{\bf f}^{\rm A}, \label{57}
\end{equation}
where
\begin{equation}
{\bf f}^{\rm AM} = -\frac{1}{2}\varepsilon_0 E^2 \bf{\nabla \varepsilon} \label{58}
\end{equation}
is common for the Abraham and Minkowski tensors and may therefore be called the Abraham-Minkowski force, acting at dielectric boundaries especially, and
\begin{equation}
{\bf f}^{\rm A} = \frac{n^2-1}{c^2}
\frac{\partial}{\partial t}{\bf (E\times H)} \label{59}
\end{equation}
is known as the Abraham term. We have here assumed a nonmagnetic medium, assumed the constitutive relation in the form ${\bf D}=\varepsilon_0 \varepsilon {\bf E}$, and omitted the electrostriction effect.

The reason why the effect of the Abraham term does not show up in optical experiments under stationary conditions, is of course that it fluctuates out when averaging over a period.  The force component left over, the Abraham-Minkowski term, is on the other hand easily detectable with modern technology.

The common way to detect the Minkowski tensor experimentally has been somewhat indirect, namely to observe the total {\it momentum flux} in an electromagnetic wave. Here, the accurate radiation pressure experiments of Jones {\it et al.} \cite{jones54,jones78} play an important role (the summary  of Jones' scientific works given in his book \cite{jones88} is interesting and highly recommended). In these experiments the radiation pressure on a mirror situated in a liquid was measured, and was found to increase proportionally to the refractive index $n$ in the liquid, in agreement with the prediction of the Minkowski tensor. To go into some detail at this point, assume that a stationary wave propagates in the $x$ direction and is reflected at the mirror having reflection coefficient $R$. Then the  surface stress $\sigma_x$ on the wall is the same as the momentum flux,
\begin{equation}
\sigma_x=\frac{n}{c}(1+R)S_x^{(i)}, \label{60}
\end{equation}
where $S_x^{(i)}$ is Poynting's vector. If we by  assume instead vacuum surroundings, (index zero), we obtain analogously $\sigma_0=(1/c)(1+R)(S_x^{(i)})_0$, taking the value of the reflection coefficient to be the same. Then assuming the same incident Poynting's vector in the two cases, $S_x^{(i)}=(S_x^{(i)})_0$, we find the simple proportionality $\sigma_x/\sigma_0=n$ that is actually observed.
This experiment was discussed more extensively in Refs.~\cite{brevik79} and \cite{brevik14}.

 A second classic experiment within the same category is the photon drag experiment of Gibson {\it et al.} \cite{gibson70}. In a semiconductor, a longitudinal electric field $E$ can be produced due to the transfer of momentum from the radiation field to the electrons in the valence or conduction bands. This field results from charges being driven down the dielectric materia. Under open-circuit conditions in a finite rod the current must be zero, and what is measured is the voltage between the two ends. Then $E$ can be determined. We abstain from going into further detail here  (a more extensive treatment is given in Ref.~\cite{brevik86}), but limit ourselves to stating that the natural interetation of the experiment is to identify the total momentum with Minkowski's momentum.

 Third, we mention the experiment of Campbell {\it et al.} \cite{campbell05} on the photon recoil momentum in a Bose-Einstein condensate (BEC). A main point here is that a BEC has a high density compared with that of laser cooled atomic cloud, thus facilitating the measurement of the refractive index-dependent recoil. In the experiment,  the recoil momentum of atoms caused by photon absorption was found to be  $\hbar k=\hbar n\omega /c$, in accordance with the Minkowski theory. Compare with the theoretical expression (\ref{23}) above, for the Minkowski momentum.

 \subsection{Remarks on a few recent experiments}

 We leave now the classic experiments listed above, and survey briefly some recent experiments on radiation pressure.

 \bigskip

 \noindent (1) The experiment of Astrath {\it et el.} \cite{astrath14}. This is a very beautiful experiment, showing the deformation of a free water surface when illuminated by a laser beam falling normally on it  from above. Both a stationary laser beam, and a pulsed one, were investigated. Under stationary conditions (optically pumped semiconductor laser, power 7 W at 532 nm) a maximum deformation of about 30 nm was observed at the center. In order to describe the deformation of the surface hydrodynamically, also the electrostriction part of the force (omitted in Eq.~(\ref{57}) above) had to be included.

Does this experiment demonstrate the existence of the Minkowski momentum? In our opinion the answer is no. All that is needed to describe the outcome of the experiment, is the Abraham-Minkowski force ${\bf f}^{\rm AM}$ defined in Eq.~(\ref{58}), common for the Abraham and Minkowski tensors, eventually  augmented by the electrostriction term if one wishes to calculate the interior pressure distribution.  There is no direct reference to the photon momentum here.

Actually, this experiment falls within the same category as the classic experiment of Ashkin and Dziedzic, also that operating in water  \cite{ashkin73}, (analyzed in detail in Ref.~\cite{brevik79}), as well as  the similar experiments considered  for instance in Refs.~\cite{casner03,hallanger05,birkeland08,wunenburger11,aanensen13} where the dielectric medium operating in the vicinity of the critical point. In that way the surface tension could be  reduced to about one millionth of that of an air-water surface. In all these cases, the force responsible for the elevation of the free surface was the term ${\bf f}^{\rm AM}$ only.

\bigskip

\noindent (2) The experiment of Zhang {\it et al.} \cite{zhang15} is another interesting recent work analyzing the deflection of a free liquid surface when illuminated by a laser beam (cf. also the Comment in Ref.~\cite{nature15}). As liquids, the experiment made use of mineral oil, and water, observing whether the boundary formed a convex defocusing surface, or a concave focusing surface. The beam was a Gaussian 1 mm wide laser, operating at 532 nm, at an angle of incidence of 3$^o$. The experiment  showed that a focusing effect took place. The experiment was supplied with detailed calculations, based  upon Gaussian beam theory. From this, the authors concluded that the experiment showed the correctness of Abraham's pressure.

In spite of the accurate experimental work involved here, we have however to conclude again that this experiment has very little to do with optical momentum. All that is needed to explain the outcome of the experiment, is the Abraham-Minkowski force ${\bf f}^{\rm AM}$, giving the gradient surface force at the air-liquid boundary. The surface pressure governing the deflection of the surface follows by integration across the boundary layer. (If one wants to describe the local pressure distribution in the liquid, one has to include the electrostriction force, which ensures that the liquid becomes deflected as a coherent whole.) This experiment is thus basically of the same kind as those we considered under item (1) above.

\bigskip
\noindent (3) As a third example we shall consider the experiment of She {\it et al.} \cite{she08}. This experiment is different from those above: a low-intensity laser beam (power 6.4 mW) was sent sent through a hanging 1.5 mm long silica glass fiber, and was found to produce a sideways deflection of the lower end, of magnitude of about 10 $\mu$m. Also, pulsed laser beams were investigated. The conclusion of the authors was they had solved  the Abraham-Minkowski problem by demonstrating that the fiber was carrying the Abraham momentum.

Here we have to object again: this experiment has most likely very   little to do with optical momentum, although the situation is not quite clear. The optical force on the medium, giving rise to the deflection, is   ${\bf f}^{\rm AM}$, acting everywhere there are geometric inhomogeneities where the permittivity is position dependent. The most actual interpretation of the experiment in our view is  that (i) there are mechanical imbalances in the fiber due to the fabrication process  leading to lack of azimuthal asymmetry, or (ii) that the lower end of the fiber is not cut precisely orthogonally, thus giving rise to a sideways  component in the surface force at this end.  In the latter case the angle of inclination need not be large to produce the mentioned   transverse deflection of 10 $\mu$m. In both cases, it is ${\bf f}^{\rm AM}$ that comes into play.

However, in this case some care has to be taken regarding the role of the Abraham force ${\bf f}^{\rm A}$. Assume that the light sent through the pulse is a short {\it pulse}. When the pulse enters the fiber from above, a downward impulse is transferred to it at its upper end. As the fiber is suspended, this does not influence the dynamics of the fiber. When the pulse exits the fiber, an upward impulse is on the other hand transferred, assuming perfect azimuthal symmetry. But it  is difficult to imagine how this should give rise to a sideways motion. We conclude that the explanation of the effect is not entirely clear; it may  depend on the detailed mechanical structure of the fiber.

We mention that critical comments on the interpretation of this experiment have been given also before, in Refs.~\cite{mansuripur09} and \cite{brevik09}

\subsection{A proposal to measure the Abraham force by use of whispering gallery modes}

Finally, we will discuss a possibility for measuring the Abraham force in optics. To our knowledge such an experiment  has so far never been done. The idea has its background in the 1975 experiments of Walker {\it et al.} \cite{walker75a,walker75b}, operating at low frequencies, allowing detection of the oscillations themselves. Thus with a dielectric cylindrical  shell  suspended in the gravitational field,  containing time varying crossed electric and magnetic strong  fields, the authors were able to detect the Abraham force (\ref{59}) directly.

In optics, where the oscillations themselves are unobservable, we can nevertheless  envisage some kind of analogy to the experiment of Walker {\it et al.} by exploiting the characteristic properties of {\it whispering gallery modes} in  microspheres. Such spheres are known to possess  large circulating powers, in excess of 100 W, close to the rim. If the sphere is suspended in the gravitational field and fed with a beam modulated at a frequency $\omega_0$, the same as the oscillation frequency of the sphere, the systems becomes to a vertical torque due to the Abraham optical force.

This idea was actually discussed in some detail in Ref.~\cite{brevik10}, and we do not consider it further here. We mention, though, that choosing a power of 100 W, and a large modulation frequency of $\omega_0 =$1000 rad/s, we obtain
\begin{equation}
N_z \sim 10^{-19}~ {\rm N~m} \label{61}
\end{equation}
as a typical value for the Abraham torque. This value is small, but might be enhanced by optimal choices for the optical and mechanical parameters.

\section{ Summary}

We have assumed a simple dielectric medium, with a real and nondispersive refractive index $n=\sqrt{\varepsilon \mu}.$ We started by demonstrating the great adaptability of Minkowski's energy-momentum tensor to the canonical formalism, both classically and quantum mechanically. This is physically related to the fact that the Minkowski tensor is divergence-free for a pure radiation field, thus leading to a four-vector property of the total energy and momentum. The fact that this four-vector becomes spacelike, so clearly demonstrated in  the Cherenkov effect, implies no difficulty in this regard. The use of our mapping technique by means of the transformation matrix (\ref{5}), made the transition between electrodynamics in vacuum and in a medium quite transparent; these two formulations are closely linked together.

From a physical viewpoint it is however not the Minkowski tensor, but instead the Abraham tensor, which appears to give the basic force \cite{landau84}. The reason why the Abraham force usually does not show up in optical experiments, is that it fluctuates out.
The critical analysis of various experiments in optics given in Sec. V, was to clear up to what extent they give information about electromagnetic momentum. Actually, in some cases experiments claiming to 'solve' the Abraham-Minkowski problem, in reality only demonstrate the action of the force ${\bf f}^{\rm AM}$, which acts in the air-surface boundary and is common for the Abraham and Minkowski tensors. Perhaps it would be possible, after all, to design an experiment in optics in which the Abraham force turns up. This was the topic of our last sub-section, discussing an optical variant of the 1975 experiments of Walker {\it et al.} \cite{walker75a,walker75b}.

\end{document}